\documentclass{article}
\usepackage{graphics,graphicx,psfrag,color,float, hyperref}
\usepackage{amsmath}
\usepackage{amssymb}
\usepackage{bm}

\usepackage{amsmath}
\usepackage{amssymb}
\usepackage{amsfonts}
\usepackage{amsthm, hyperref}
\usepackage {times}
\usepackage{amsfonts}

\newcommand{\beq}{\begin{equation}}
\newcommand{\enq}{\end{equation}}
\newcommand{\bel}{\begin{lemma}}
\newcommand{\enl}{\end{lemma}}
\newcommand{\bet}{\begin{theorem}}
\newcommand{\ent}{\end{theorem}}

\newcommand{\tr}{\mathrm{Tr}}
\newcommand{\vr}{\mathrm{Var}}

\newcommand{\rv}[1]{\mathbf{#1}}

\newcommand{\E}{\mathbb{E}}

\newcommand{\ceil}[1]{\left\lceil #1 \right\rceil}

\newcommand{\eps}{\varepsilon}
\newcommand*{\cC}{\mathcal{C}}
\newcommand*{\cA}{\mathcal{A}}
\newcommand*{\cH}{\mathcal{H}}
\newcommand*{\cM}{\mathcal{M}}

\newcommand*{\cG}{\mathcal{G}}

\newcommand*{\cN}{\mathcal{N}}

\newcommand*{\cT}{\mathcal{T}}
\newcommand*{\cX}{\mathcal{X}}

\newcommand*{\cZ}{\mathcal{Z}}
\newcommand*{\cE}{\mathcal{E}}
\newcommand*{\cU}{\mathcal{U}}

\newcommand*{\cV}{\mathcal{V}}
\newcommand*{\cY}{\mathcal{Y}}

\newcommand{\bra}[1]{\langle #1|}
\newcommand{\ket}[1]{|#1 \rangle}

\mathchardef\mhyphen="2D

\newcommand*{\renyi}{R\'{e}nyi }

\mathchardef\mhyphen="2D

\newtheorem{definition}{Definition}
\newtheorem{claim}{Claim}

\newtheorem{comment}{Remark}
\newtheorem{theorem}{Theorem}
\newtheorem{lemma}{Lemma}

\begin {document}
\title{One-shot Marton inner bound for classical-quantum broadcast channel}
\author{Jaikumar Radhakrishnan${}^*$ 
\and 
Pranab Sen${}^*$ 
\and 
Naqueeb Warsi\thanks{
School of Technology and Computer Science, Tata Institute of Fundamental Research,
Mumbai 400005, India.
Email: 
{\sf 
\{jaikumar,naqueeb\}@tifr.res.in, pgdsen@tcs.tifr.res.in
}
}
}
\date{}

\maketitle
\begin{abstract}
We consider the problem of communication over a classical-quantum
broadcast channel with one sender and two receivers. Generalizing the
classical inner bounds shown by Marton and the recent quantum
asymptotic version shown by Savov and Wilde, we obtain one-shot inner
bounds in the quantum setting. Our bounds are stated in terms of
smooth min and max \renyi divergences. We obtain these results using a
different analysis of the random codebook argument and employ a new
one-shot classical mutual covering argument based on rejection
sampling.  These results give a full justification of the claims of
Savov and Wilde in the classical-quantum asymptotic iid setting; the techniques
also yield similar bounds in the information spectrum setting.
\end{abstract}
\maketitle
\section{Introduction}
We consider the problem of communication over a broadcast channel with
one sender (Alice) and two receivers (Bob and Charlie).  They have
access to a channel that takes one input $X$ (supplied by Alice) and
produces two outputs $Y$ and $Z$, received by Bob and Charlie
respectively. The characteristics of the channel are given by $p(y,z
\mid x)$ . The goal is to obtain bounds on the rates at which Alice
may transmit messages simultaneously to Bob and Charlie. 

\begin{figure}[H]
\centering
\vspace{0.4cm}
\resizebox{0.9\textwidth}{!}{
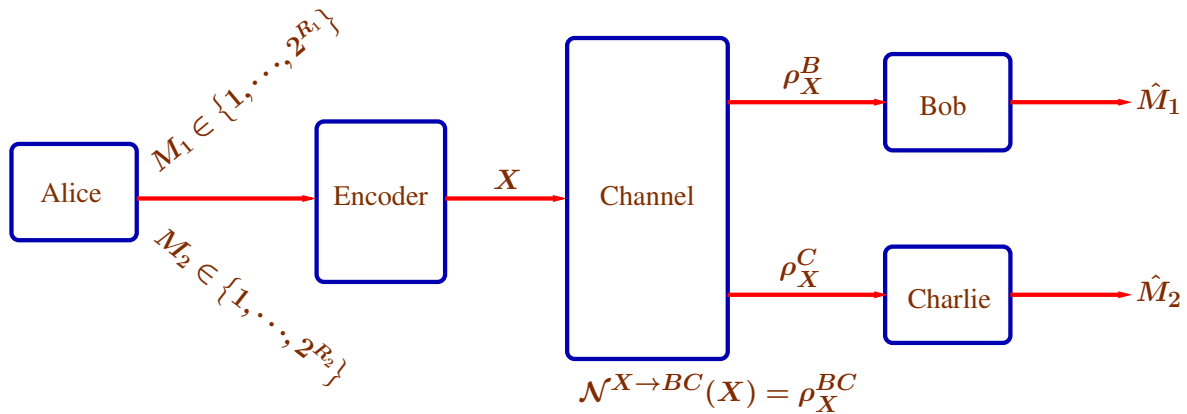
}
\vspace{0.4cm}
\caption{Broadcast channel communication scheme.}
\label{Broadcast channel}
\end{figure}

\paragraph{Marton bound:} An achievable rate 
region for this channel was given by Marton~\cite{marton-79}, in the asymptotic iid setting
who showed the following.
\begin{theorem}
\label{mar}
Fix a discrete memoryless broadcast channel given by $p(y,z\mid x)$.
Let a pair of random variables $(U,V)$ taking values in $\cU \times
\cV$ and a function $f: \cU \times \cV \rightarrow \cX$ be given;
suppose the random variables $(U,V,Y,Z)$ have joint probability mass
function $p(u,v,y,z) = p(u,v)p(y,z \mid f(u,v))$.  Let
$(R_1,R_2)$ be such that
\begin{align}
R_1 &< I[U;Y],\\
R_2 &< I[V;Z],\\
\label{sumra}
R_1+R_2 & < I[U;Y]+I[V;Z]-I[U;V].
\end{align}
Then, the rate pair $(R_1,R_2)$ is achievable.
 \end{theorem}
A quantum version of the broadcast channel was considered by Savov and
Wilde~\cite{savov-wilde-isit}, where instead of $p(y,z \mid x)$,
the channel is characterized by density matrices $\rho_x^{BC}$ (note
that the channel takes classical input, so $x$ is classical).  A communication scheme over a classical-quantum broadcast channel is illustrated in Figure \ref{Broadcast channel}.
Savov and Wilde \cite{savov-wilde-isit} formulated the following quantum version of the Marton inner bound in the asymptotic iid setting.
\begin{theorem} \label{savov-wilde}
Let $(\cX,\cN:x \mapsto \rho^{BC}_x)$ be a classical-quantum broadcast
channel. Let a pair of random variables $(U,V)$ taking values
in $\cU \times \cV$ and a function $f: \cU \times \cV \rightarrow \cX$
 be given; consider the state
\beq
\rho^{UVBC} = \sum_{(u,v)\in
  \cU\times\cV}p_{UV}(u,v)\ket{u}\bra{u}^U\otimes
\ket{v}\bra{v}^V\otimes\rho^{BC}_{f(u,v)}.  \nonumber
\enq 
Let $(R_1,R_2)$ be such that
\begin{align*}
R_1 &< I[U;B],\\
R_2 &< I[V;C],\\
R_1+R_2 &< I[U;B]+I[V;C] - I[U;V],
\end{align*}
where the information theoretic quantities above are computed with
respect to the state $\rho^{UVBC}$.  Then, the rate pair $(R_1,R_2)$ is
achievable (see \cite{savov-wilde-arxiv-v3} for the definition of achievable rate pair). 
\end{theorem}

\paragraph{Motivation:} Our work is motivated by the work of Savov and
Wilde~\cite{savov-wilde-isit} mentioned above, who base their proof on
the presentation of Marton's bound in the book of El Gamal and Kim
\cite{Gamal-Kim-book}. The argument proceeds as follows. The sender
uses a randomly generated codebook. The codewords to be fed into the
channel are arranged in a rectangular array. The rows are partitioned
into $2^{nR_1}$ bands and the columns into $2^{nR_2}$ bands. There is
one band of rows for each message $m_1$ that Alice might need to send
to Bob, and one band of columns for each message $m_2$ that Alice
might need to send to Charlie. On receiving $(m_1,m_2)$, Alice picks a
codeword from the intersection of the corresponding bands and feeds it
into the channel. Bob and Charlie, on receiving their share of the
channel output, try to determine the intended messages $m_1$ and
$m_2$, that is, locate the corresponding row and column bands.  El
Gamal and Kim show that with high probability the correct bands can be
identified by Bob and Charlie.  Formally, this is done by applying the
union bound to upper bound the probability of decoding a wrong
band. This part of the proof is not straightforward to translate into
the quantum setting. In fact, the argument presented by Savov and Wilde ~\cite{savov-wilde-isit} leaves a gap;
in a subsequent version
of their paper~\cite{savov-wilde-arxiv-v2}, this gap is acknowledged, but the old gaps are
still present in the new version. Very recently, Savov and Wilde
addressed this gap again in an update of their archive
paper~\cite{savov-wilde-arxiv-v3}, and borrowing several ideas
from an earlier version of our work
~\cite{radhakrishnan-sen-warsi-archive-v1}, provide a complete
justification of their claims in the asymptotic iid setting.

\paragraph{Our results:} We consider the above problem 
of communication over a classical-quantum channel in the one-shot
setting. We show the following version of Marton bound.
\begin{theorem}
\label{martonquantum-oneshot}
Let $(\cX,\cN:x \mapsto \rho^{BC}_x)$ be a classical-quantum broadcast
channel. Let a pair of random variables is $(U,V)$ taking values in
$\cU \times \cV$ and a function $f: \cU \times \cV \rightarrow \cX$ be
given; consider the state \beq
\label{jointstatem}
\rho^{UVBC} = \sum_{(u,v)\in
  \cU\times\cV}p_{UV}(u,v)\ket{u}\bra{u}^U\otimes
\ket{v}\bra{v}^V\otimes\rho^{BC}_{f(u,v)}.  \enq 
Let $(R_1,R_2)$, $\eps$, $\eps_\infty$, $\eps_0$ and $\tilde{\eps}$ 
be such that
\begin{align}
\label{a1}
R_1&\leq I^{{\eps_0}}_{0}[U;B] - 5 \log\frac{1}{\tilde{\eps}} - 2\\
\label{a2}
R_2&\leq I^{{\eps_0}}_{0}[V;C] - 5 \log\frac{1}{\tilde{\eps}} - 2\\
\label{a3}
R_1+R_2 & \leq I^{{\eps_0}}_{0}[U;B] + I^{{\eps_0}}_{0}[V;C]-I^{\eps_{\infty}}_{\infty}[U;V] -11\log\frac{1}{\tilde{\eps}}-5,
\end{align}
where ${\eps}_{\infty} \leq \frac{1}{4}$ and $40\tilde{\eps}+16\eps_0
\leq \eps$. Then, there exists a $(R_1,R_2,\eps)$-classical-quantum
broadcast channel code. The information theoretic quantities mentioned
in \eqref{a1}, \eqref{a2} and \eqref{a3} are calculated with respect
to the classical-quantum state given in
(\ref{jointstatem}). (Classical-quantum broadcast channel code is defined in
Section~\ref{sec:classical-quantum} and $I^{{\eps_0}}_{0}$ and $I^{\eps_{\infty}}_{\infty}$ are defined in Definition \ref{csmoothmindiv} and Definition \ref{smoothorderinf} respectively.)
\end{theorem}
This result implies the result of Savov and
Wilde~\cite{savov-wilde-arxiv-v3}.  Our method also yields the following
one-shot version of Marton's inner bound in the classical setting.
\newcommand{\zeroepsilon}{\epsilon_0}

\newcommand{\thirdepsilon}{\tilde{\epsilon}}
\newcommand{\ty}{\tilde{y}} \newcommand{\tz}{\tilde{z}}
\newcommand{\tm}{\tilde{m}} \newcommand{\rvI}{\mathbf{I}}
\newcommand{\rvJ}{\mathbf{J}} \newcommand{\indI}{\mathbb{I}}
\newcommand{\event}{\mathcal{E}}
\begin{theorem} \label{thm:classical}
Consider a classical broadcast channel given by $p(yz \mid x)$, where
$x \in \cX$, $y \in \cY$ and $z \in \cZ$.  Suppose there is a pair of
random variables $(U,V) \in \cU \times \cV$ and a function $f: \cU
\times \cV \rightarrow \cX$. Let $(Y,Z)$ be random variables such that
$\Pr\{(Y,Z)= (y,z) \mid U=u,V=v\} = p(y,z \mid f(u,v))$. Further,
suppose $(R_1,R_2)$ and $\eps$, $\eps_0$, $\eps_\infty$, and
$\thirdepsilon$ are such that
\begin{eqnarray}
R_1 &\leq& I_0^{\eps_0}[U;Y] - 5\log \frac{1}{\tilde{\eps}} -2 \\
R_2 &\leq& I_0^{\eps_0}[V;Z] -5\log \frac{1}{\tilde{\eps}} -2 \\
R_1 + R_2 &\leq& I_0^{\eps_0}[U;Y] +  I_0^{\eps_0}[V;Z]
               -I_\infty^{\eps_\infty}[U;V] - 11 \log \frac{1}{\tilde{\eps}} -5,
\end{eqnarray}
where ${\eps}_{\infty} \leq \frac{1}{4}$ and $37\tilde{\eps}+ 8
\eps_0 \leq \eps$. Then, there is a one-shot
$(R_1,R_2,\eps)$-classical broadcast code for the channel.
(Classical broadcast channel code is defined in Section~\ref{sec:classical} and $I_{0}^{\eps_0}$ and $I_\infty^{\eps_\infty}$ are defined in Definition \ref{mindiv} and Definition \ref{smoothorderinf} respectively.)
\end{theorem}

\paragraph{Techniques:} Our proof follows along the lines of the proof in 
El Gamal and Kim~\cite{Gamal-Kim-book} for the classical Marton
bound. As before, we generate a rectangular array, whose rows and
column indices are chosen independently according to the marginal
distributions of $U$ and $V$; furthermore, as in the original proof,
we partition the rows and columns into bands of appropriate sizes.
There are two major difficulties that one encounters.
\begin{enumerate}
\item[(a)] First, given a message pair $(m_1,m_2)$, we do not have a
  natural analogue of joint typicality to help us choose a $(u,v)$
  pair from the subcodebook.  Furthermore, in the iid setting it is
  well-known that if a jointly typical pair is used as input to the
  channel, the output is very likely to be jointly typical with the
  input; however, we cannot exploit such facts in the one-shot
  setting. Instead, we use rejection sampling to ensure that the
  resulting probability distribution is very close to the ideal joint
  distribution on $(U,V)$ and the outputs for the channel.

\item[(b)] Second, the difficulty mentioned above with applying the
  union bound, particular to the asymptotic quantum setting, are
  present in the one-shot setting as well, and equally hard to
  overcome. The solution is somewhat technical, we observe that if the
  sizes of the bands are tuned carefully, we have the liberty to
  overestimate the probability of error, and obtain a good bound.
  {\em Our analysis technique in fact shows that if the receivers
    employ a standard pretty good measurement technique, then they not
    only decode the transmitted bands correctly, but also recover the
    row and column index pair that was used for transmission.}  This
  analysis of error at the two receivers differs from the standard
  analysis in the normal point to point classical-quantum channel
  coding problem, as well as from the `decoding up to the band'
  arguments of El Gamal and Kim.

\end{enumerate}
The new methods seem necessary to deal with the additional
complications that arise in the one-shot setting.
\begin{comment}
As mentioned earlier the proof of Theorem \ref{mar} relies on joint typicality encoding at the transmitter end. As a result of this encoding scheme there is a loss in the sum transmission rate. This loss is reflected by the term $I[U;V]$ in \eqref{sumra}. Likewise, in the one-shot setting this loss in the sum transmission rate is reflected by the term $-5-I_{\infty}^{\eps_{\infty}}[U;V]$ in \eqref{a3}. We will argue in Remark \ref{LbI} that for $\eps_{\infty} \leq \frac{1}{4}$, we have $I_{\infty}^{\eps_{\infty}} [U;V]>-1.$ Thus, implying that $5+I_{\infty}^{\eps_{\infty}}[U;V]>0$.
\end{comment}
\paragraph{Related work:}
Prior to Savov and Wilde \cite{savov-wilde-isit},
Yard et al.~\cite{yard-2011} proved
superposition coding inner bounds for classical and quantum
communication over a quantum broadcast channel. In their paper
\cite{dupuis-broadcast-2010}, Dupuis, Hayden and Li prove 
Marton's inner
bound for transmission of quantum information over an entanglement
assisted quantum-quantum broadcast channel using decoupling techniques.
They also prove one shot bounds for the same problem in terms of
certain conditional entropies.
It is not clear if their results or techniques imply anything for
sending classical information. 
On the other hand, our
work does not imply anything about sending quantum information. Thus
their work and ours are best thought of 
as incomparable. Dupuis, Hayden and Li derive their analog of Marton's
bound for the asymptotic iid case by applying a quantum asymptotic 
equipartition theorem to their conditional entropy based one shot 
results. Our one shot bounds are stated not in terms of conditional
entropies but in terms of two
fundamental smooth Renyi divergences.
In particular, they allow us to easily prove an analog of Marton's 
bound for the asymptotic non-iid (information spectrum) scenario, which
does not seem possible with conditional entropies.

A one-shot inner bound for classical broadcast channels was also proved 
by Verd\'{u} (see \cite[Theorem 8]{verdu-nonasymptotic}).  Stated using 
the terminology above, 
Verdu's inequalities can be stated as follows.
\begin{align*}
R_1 &\leq I_0^{\eps_0}[U; Y]  - \ln \frac{1}{\gamma}\\
R_2 & \leq  I_0^{\eps_0}[V; Z] - I_\infty^{\eps_\infty}[U; V] - 
         2 \ln\frac{1}{\gamma}\\
\eps & \leq 2 \eps_0 + \eps_\infty+ 2\gamma + \exp(- \frac{1}{\gamma}).
\end{align*}
Note that apart from the dependence on $\eps$, our inner bound region
is strictly larger than Verd\'{u}'s. Furthermore, while the strategy 
employed by
Verd\'{u} allows the decoding of the transmitted bands correctly with
high probability, we achieve more by decoding the actual row and
column.

\begin{comment}
The above achievability region is slightly bigger than what is claimed 
in \cite[Theorem 8]{verdu-nonasymptotic} because 
$I^{\eps_{0}}_{0}[U;Y]$ and $I^{\eps_{0}}_{0}[V;Z]$ we use is somewhat bigger than the corresponding quantity in Verd\'{u}'s formulation; however, the difference is very minor.
\end{comment}

The technical difficulty in ensuring unique decoding in several
classical settings related to ours has been recognised and addressed
in several recent works~\cite{vinod-nonunique, lapidoth-unique,
  minero-nonuinque, grover-nonunique}. In particular, Minero, Lim and
Kim~\cite[Lemma 1]{minero-nonuinque} achieve unique decoding for the
rate regions associated with the Gelfand-Pinsker bound in the
asymptotic setting by controlling the perturbations in distributions
caused by conditioning on other events. Their analysis makes critical
use of the asymptotic equipartition property (AEP) available in the
asymptotic iid setting. We, working in the one-shot and non-iid
setting do not have recourse to such tools. Instead, we observe that
by carefully controlling the band sizes while generating the code, one
can simply {\em over count} and bound the probability of error. We believe
this method is applicable to other settings as well (see
Remark~\ref{rem:overcounting} below). Furthermore, this method of
analysis works in the quantum setting with almost no change.

\paragraph{Asymptotic iid and non-iid bounds.} Our bounds imply the bounds 
obtained earlier for the same problem in the iid setting. The
asymptotic information spectrum setting pioneered by Han and
V\'{e}rdu~\cite{han-verdu-93} and its quantum version due to Hayashi
and Nagaoka~\cite{Hayashi-noniid} allows one to derive meaningful
bounds on rates even in the absence of the iid assumption; however,
the analysis is often more challenging in these settings. The bounds
in our work are expressed using smooth \renyi quantities. The close
relationship between these quantities and the quantities that typically
arise in the information spectrum setting (see Datta and
Renner~\cite{datta-renner-2009}) allows us to conclude similar
bounds in the asymptotic case, in both the non-iid (information
spectrum) and iid setting.

%

\section{The classical-quantum one-shot bound}
\label{sec:classical-quantum}
\begin{definition}[Channel]
Let $\cX$ be a finite alphabet. We model a classical-quantum broadcast
channel between parties Alice, Bob and Charlie as a map 
\beq \cN: x
\mapsto \rho_x^{BC}, 
\enq 
where $x \in \cX$ is the input given to
the channel by Alice, and $\rho_x^{BC}$ is the joint state of Bob and
Charlie in the Hilbert space $\cH_A \times \cH_B$. The resulting state
of Bob is then modelled as $ \rho_x^B = \tr_C \rho_x^{BC}$, and the
state of Charlie is modelled as $\rho_x^C = \tr_B \rho_x^{BC}$.
\end{definition}
Our goal is to use this channel to enable Alice to transmit a pair of
messages $(m_1,m_2) \in \cM_1 \times \cM_2$ (for some large sets
$\cM_1$ and $\cM_2$) such that Bob can recover $m_1$ and Charlie can
recover $m_2$.
\begin{definition}[Encoding, Decoding, Error]
An $(R_1, R_2, \eps)$-classical-quantum broadcast channel code consists of 
\begin{itemize}
\item an encoding function $F: [2^{R_1}] \times [2^{R_2}] \rightarrow \cX$, and
\item two decoding POVMs $\{\cT^B_{m_1}: m_1 \in [2^{R_1}]\}$ and 
 $\{\cT^C_{m_2}: m_2 \in [2^{R_2}]\}$ such that the average probability
of error
\beq
\frac{1}{2^{R_1+R_2}}\sum_{(m_1,m_2)\in [2^{R_1}]\times [2^{R_2}]}
p_e(m_1,m_2) \leq \eps,
\enq
where 
\[ p_e(m_1,m_2) = \tr\left[\left(\mathbb{I} - \cT^B_{m_1} \otimes \cT^C_{m_2}\right)
                 \cN(F(m_1,m_2))\right],\] is the probability of error when
                 Alice uses this scheme to transmit the message pair 
$(m_1,m_2)$.
\end{itemize}
\end{definition}
Our one-shot version of the Marton inner bound will be stated
in terms of min and max \renyi divergences which are defined as follows.
\begin{definition}(Smooth quantum min \renyi divergence \cite{wang-renner-prl})
\label{csmoothmindiv}
Let $\rho^{UB}:=\sum_{u\in\cU}p_{U}(u)\ket{u}\bra{u}^U\otimes\rho_u^B$ be a classical quantum state. For $\eps\in[0,1)$ the smooth min \renyi divergence between the systems $U$ and $B$ denoted is
\beq
I^{\eps}_{0}[U;B]:=\sup_{\substack{0\preceq\Gamma^{UB}\preceq\mathbb{I}\\ \tr\left[\Gamma^{UB}\rho^{UB}\right]> 1-\eps}}-\log\tr\left[\Gamma^{UB}\left(\rho^{U}\otimes\rho^B\right)\right]. \nonumber
\enq
\end{definition}
\begin{definition}(Smooth max \renyi divergence \cite{warsi-isit-2013})
\label{smoothorderinf}
For random variables $(U,V)\sim p_{UV}$ with range  $\cU\times\cV$
and $\eps \in [0,1)$ we have
\beq
I^\eps_{\infty}[U;V]=\inf_{\substack{\cG \subseteq \cU\times\cV\\p_{UV}(\cG)> 1-\eps}}
\sup_{(u,v) \in \cG} \log \frac{p_{UV}(u,v)}{p_U(u)p_V(v)}. \nonumber
\enq
\end{definition}
\begin{comment}
\label{LbI}
We note here that for $\eps \in \left[0, \frac{1}{2}\right]$ we have $I^\eps_{\infty}[U;V]> -1$. To verify this let $$\cA:=\left\{x: p_{UV}(u,v) \leq p_U(u)p_V(v)2^{I^\eps_{\infty}[U;V]}\right\}.$$ Thus, from the definition of the set $\cA$ we conclude that $\sum_{(u,v) \in \cA}p_{UV}(u,v) > 1- \eps$. It now further follows from the definition of $\cA$ that $2^{I^\eps_{\infty}[U;V]}\sum_{(u,v) \in \cA}p_{U}(u)p_V(v) > 1- \eps$. Our observation now immediately follows from this.
\end{comment}

We are now ready to prove Theorem~\ref{martonquantum-oneshot}.

\subsection{Proof of Theorem \ref{martonquantum-oneshot}}
\label{q1}
We need to describe the encoding function $F: [2^{R_1}] \times
[2^{R_2}] \rightarrow \cX$ and suitable POVMs that will be used for
decoding.  We will adapt the scheme suggested by Marton as presented
in El Gamal and Kim~\cite{Gamal-Kim-book}, to the quantum one-shot
setting.

Let $\rho^{UVBC}$ and $f$ be as in the
statement of the theorem, and $(R_1,R_2)$ satisfy the required
inequalities.  In the following we set
\begin{eqnarray*}
I_\infty  & = & I^{\eps_{\infty}}_\infty[U;V]; \\
I_0^B     & = & I^{\eps_0}_0[U;B];\\
I_0^C     & = & I^{\eps_0}_0[V;C].\\
\end{eqnarray*}
Let $\Gamma^{UB}$ be such that $\tr\left[\Gamma^{UB}\rho^{UB}\right]\geq 1 -
\eps_0$ and $\tr\left[\Gamma^{UB}\left(\rho^{U}\otimes\rho^B\right)\right]=
2^{-I_0^B}$. Similarly, let $\Gamma^{VC}$ be such that $\tr\left[\Gamma^{VC}\rho^{VC}\right]\geq 1 -
\eps_0$ and $\tr\left[\Gamma^{VC}\left(\rho^{V}\otimes\rho^C\right)\right]=
2^{-I_0^C}$ where $\rho^{UB} = \tr_{VC}\left[\rho^{UVBC}\right]$; $\rho^{U} = \tr_{VBC}\left[\rho^{UVBC}\right]$; $\rho^{B} = \tr_{UVC}\left[\rho^{UVBC}\right]$; $\rho^{VC} = \tr_{UB}\left[\rho^{UVBC}\right]$; $\rho^{V} = \tr_{UBC}\left[\rho^{UVBC}\right]$ and $\rho^{C} = \tr_{UVB}\left[\rho^{UVBC}\right]$. Choose positive integers $r_1$ and $r_2$ such that
\begin{eqnarray}
 R_1 + r_1 &\leq& {I_0^B-4\log\frac{1}{\tilde{\eps}}-1}; \label{eq:rfirstq} \\
 R_2 + r_2 &\leq& {I_0^C - 4\log\frac{1}{\tilde{\eps}}-1};\label{eq:r1firstq} \\
r_1, r_2 & \geq & \log \frac{1}{\tilde{\eps}}; \label{eq:rlastq1}\\
r_1 + r_2 & = & \ceil{I _\infty + 3\log \frac{1}{\tilde{\eps}}}\label{eq:r2firstq}. 
\end{eqnarray}
\paragraph{Verification for the existence of $(r_1,r_2)$ satisfying (\ref{eq:rfirstq})--(\ref{eq:r2firstq}) :}  To see that such a choice exists we may, e.g., start with $r_1, r_2 = \ceil{\log \frac{1}{\tilde{\eps}}}.$ Then, (\ref{eq:rfirstq})--(\ref{eq:rlastq1}) follow immediately using \eqref{a1} and \eqref{a2}. Now consider (\ref{eq:r2firstq}). Since  $\eps_{\infty} \leq \frac{1}{4},$ we have  $I_{\infty}> -1$ (see Remark \ref{LbI}). Furthermore, since $\tilde{\eps} \leq \frac{1}{40},$ we have $3\log\frac{1}{\tilde{\eps}} \geq 2 \log\frac{1}{\tilde{\eps}}+3.$ Thus, RHS $\geq \ceil{-1+3\log \frac{1}{\tilde{\eps}}} \geq 2 \ceil{\log \frac{1}{\tilde{\eps}}} =$ LHS. Now, if necessary we increase $r_1$ and $r_2$ without violating (\ref{eq:rfirstq}) or (\ref{eq:r1firstq}), until (\ref{eq:r2firstq}) is satisfied. To see that we will succeed in this, suppose at some point 
\beq
\label{contass}
r_1 + r_2  <  \ceil{I _\infty + 3\log \frac{1}{\tilde{\eps}}},
\enq
 and yet $r_1$ and $r_2$ have reached their maximum values permissible in (\ref{eq:rfirstq}) and (\ref{eq:r1firstq}), so that 
\begin{align*}
r_1 &\geq {I_0^B-R_1 - 4\log\frac{1}{\tilde{\eps}}-2};\\
r_2 & \geq {I_0^C - R_2 -4\log\frac{1}{\tilde{\eps}}-2}.
\end{align*}
But then,
\begin{align*}
r_1 + r_2 & \geq I_0^B + I_0^C- 8\log\frac{1}{\tilde{\eps}}-4 -(R_1+R_2)\\
 &\geq I_0^B + I_0^C- 8\log\frac{1}{\tilde{\eps}}-4-\left(I_0^B + I_0^C - I_{\infty} -11\log\frac{1}{\tilde{\eps}}-5\right)~~~~~~~(\mbox{using} ~~\eqref{a3})\\
 & \geq I_{\infty} + 3 \log\frac{1}{\tilde{\eps}} +1 ,
\end{align*}
contradicting our assumption \eqref{contass}.

\paragraph{The random codebook:} Let $U[1]$, $U[2]$,$\ldots, U[2^{R_1+r_1}]$ be drawn independently
according to the distribution of $U$; similarly, let $V[1]$,
$V[2]$,$\ldots, V[2^{R_2+r_2}]$ be drawn according the distribution of
$V$. These samples will be associated with rows and columns of a
$2^{R_1+r_1} \times 2^{R_2+r_2}$ matrix $\cC$, whose entries will be
elements of $\cX \cup \{\star\}$. The entry $\cC[k,\ell]$ will be
determined as follows.

For each pair $(k,\ell)$, let $\eta(k,\ell)$ be chosen independently and
uniformly from $[0,1]$. Let $\indI(k,\ell)$ be the $0$-$1$ indicator
random variable defined by
\beq
\label{jtyp}
 \mathbf{I}(k,\ell) = \mathbb{I}\left\{ \eta(k,\ell) \leq
\frac{p(U[k],V[\ell])}{2^{I_{\infty}}{ p(U[k]) p(V[\ell])}} \right\}.
\enq
Then, $\cC[k,\ell] = f(U[k], V[\ell])$ if $\mathbf{I}(k,\ell) = 1$, and
$\cC[k,\ell]=\star$ otherwise. Thus, $\cC$ is a random matrix of entries,
determined by the random choices of $(U[k], V[\ell], \eta(k,\ell))$
for $k=1,2,\ldots, 2^{R_1+r_1}$ and $\ell = 1,2, \ldots, 2^{R_2+r_2}$;
we will call this (the random matrix, together with all the associated
random choices $U[k]$, $V[\ell]$ and $\eta(k,\ell)$) the random
codebook $\cC$. Later we will fix one realization of $\cC$.

Our encoding function $F: [2^{R_1}] \times [2^{R_2}] \rightarrow \cX$
will be based on $\cC$. We partition the row indices of $\cC$ into
$2^{R_1}$ classes each with $2^{r_1}$ elements; let the $i$-th class
$\cC_1(i) = \{(i-1)2^{r_1}+1, (i-1)2^{r_1}+ 2, \ldots, i
2^{r_1}\}$. Similarly, we partition the column indices into $2^{R_2}$
classes, where the $j$-th class $\cC_2(j) = \{(j-1)2^{r_2}+1,
(j-1)2^{r_2}+ 2, \ldots, j 2^{r_2}\}$.  $F(m_1,m_2)$ will be set to
$\cC[k,\ell] \neq \star$ for some $(k,\ell) \in \cC_1(m_1) \times
\cC_2(m_2)$. However, we must ensure that the choice $(k,\ell)$ aids the
decoding process.

Below, we will see that the POVMs used by Bob and Charlie will be based on 
operators defined as follows.
\begin{eqnarray}
\label{lambu}
\Lambda^B_{u}&:=& \tr_{U}\left[\Gamma^{UB}\left(\ket{u}\bra{u}\otimes \mathbb{I}\right)\right] \\
\Lambda^C_{v} &:=& \tr_{V}\left[\Gamma^{VC}\left(\ket{v}\bra{v}\otimes \mathbb{I}\right)\right]. 
\end{eqnarray}
Similar operator was used by Wang and Renner in \cite{wang-renner-prl} to design the decoding POVM elements for finding one-shot achievable rate for the point to point classical-quantum channels.
Our choice of $(k,\ell)$ will be guided by these operators.
If $\cC[i,j] = x \neq \star$, then let
\begin{eqnarray*}
\alpha(i,j) &=& \tr\left[\Lambda^B_{U[i]}\rho^B_{f\left(U[i],V[j]\right)}\right];\\
\beta(i,j) &=&  \tr\left[\Lambda^C_{V[j]}\rho^C_{f\left(U[i],V[j]\right)}\right].
\end{eqnarray*}
If $\cC[i,j]=\star$, let $\alpha(i,j), \beta(i,j)= - \infty$.  For a
pair of messages $(m_1, m_2)$, let $F(m_1,m_2) = \cC[i,j]$, where $(i,j)
\in \cC_1(m_1) \times \cC_2(m_2)$ is the lexicographically first pair such
that $\alpha(i,j), \beta({i,j}) > 1 - 4\eps_0$; if no such $(i,j)$
exists, then let $F(m_1,m_2)$ be the first element of $\cX$.

\paragraph{Decoding:} We first consider Bob's strategy for recovering
$m_1$ on receiving the channel output $\sigma^B$. Fix a codebook.  For
each $k \in [2^{R_1+r_1}]$, we have the operator $\Lambda^B_{U[k]}$
defined above. Bob will {\em normalize} these operators, to obtain a
POVM. The POVM element corresponding to $k$ will be 
\beq \label{bobdec}
\cT^B_{k} =
\left(\sum_{k^\prime\in
  [2^{R_1+r_1]}}\Lambda^B_{U[k']}\right)^{-\frac{1}{2}}\Lambda^B_{U[k]}\left(\sum_{k^\prime\in
  [2^{R_1+r_1]}}\Lambda^B_{U[k']}\right)^{-\frac{1}{2}}. \enq 
Bob measures his state using these operators to
obtain an index $\tilde{k} \in [2^{r_1+R_1}]$ (we would like this to
be $k$, the row index used by Alice). He outputs $\tilde{m}_1$ if
$\tilde{k} \in \cC_1({\tilde{m}_1})$. Similarly, for every $\ell \in [2^{R_2+r_2}]$ Charlie has the
following POVM element 
\beq \label{chardec}
\cT^C_{\ell} = \left(\sum_{\ell^\prime\in
  [2^{R_2+r_2]}}\Lambda^C_{V[\ell']}\right)^{-\frac{1}{2}}\Lambda^C_{V[\ell]}\left(\sum_{\ell^\prime\in
  [2^{R_2+r_2]}}\Lambda^C_{V[\ell']}\right)^{-\frac{1}{2}}.  \enq Using
this POVM, Charlie measures his state $\sigma^C$ to obtain a column
index $\tilde{\ell} \in [2^{r_2+ R_2}]$, and outputs $\tilde{m}_{2}$
if $\tilde{\ell} \in \cC_{2}({\tilde{m}_2})$.

\paragraph{Joint typicality versus rejection sampling:} In the standard
argument~\cite{{Gamal-Kim-book}}, the indicator random variable
$\mathbf{I}(k,\ell)$ stands for joint typicality of $U[k]$ and
$V[\ell]$. The rejection sampling based on $I_\infty$, has the same
effect. We list below some of its properties.
\begin{enumerate}
\item[(P1)] $\E\{\mathbf{I}(k,\ell)\} \geq (1-\eps_\infty) 2^{-I_\infty}$. This property can be proven as follows.
\begin{align*}\E\{\mathbf{I}(k,\ell)\} &= \sum_{(u,v): \frac{p_{UV}(uv)}{p_U(u)p_V(v)} \leq 2^{-I_\infty}}\Pr\left\{U(k)=u\right\}\Pr\left\{V(l)=v\right\}\\& = 2^{-I_\infty}\sum_{(u,v): \frac{p_{UV}(uv)}{p_U(u)p_V(v)} \leq 2^{-I_\infty}}p_{UV}(u,v)\\
&\geq (1-\eps_\infty) 2^{-I_\infty},\end{align*}
where the last inequality follows from the definition of $I_\infty$.
\item[(P2)] For all $u$ and $v$,
\begin{eqnarray}
\E\{\mathbf{I}(k,\ell) \mid U[k]=u\} 
&\leq & \sum_v p(v) \frac{p(u,v)}{2^{I_\infty} p(u)p(v)} \leq 2^{-I_\infty}; \\
\E\{\mathbf{I}(k,\ell) \mid V[\ell]=v\}  & \leq & 
\sum_u p(u)  \frac{p(u,v)}{2^{I_\infty} p(u)p(v)} \leq
2^{-I_\infty};\\
\E\{\mathbf{I}(k,\ell)\} & \leq & 2^{-I_\infty}.
\end{eqnarray}
\item[(P3)] If $\ell \neq \ell'$, then $\mathbf{I}(k,\ell)$ and $\mathbf{I}(k,\ell')$
  are conditionally independent given $U[k]$; if $k \neq k'$, then
  $\mathbf{I}(k,\ell)$ and $\mathbf{I}(k',\ell)$ are conditionally independent
  given $V[\ell]$.

\item[(P4)] If $k \neq k'$ and $\ell \neq \ell'$, then 
 $\mathbf{I}(k,\ell)$ and $\mathbf{I}(k',\ell)$ are independent.
\end{enumerate}

\paragraph{Probability of error:} Suppose a pair of messages 
$(m_1,m_2) \in [2^{R_1}] \times [2^{R_2}]$ is transmitted by
Alice using the above scheme and is decoded as $(\tilde{m}_1,
\tilde{m}_2)$ by Bob and Charlie.  We wish to show that the
probability (averaged over the choice of the codebook) that
$(\tilde{m}_1,\tilde{m}_2) \neq (m_1,m_2)$ is at most $\eps$.  By
the symmetry in the generation of the code book, it is enough to prove
this claim for $(m_1,m_2)=(1,1)$.  There are several sources of error:
(i) Alice finds no suitable pair $(k,\ell) \in \cC_1(1) \times \cC_2(1)$;
(ii) Alice finds a suitable pair, say $(k^*, \ell^*)$, but
Bob's measurement is unable to determine the index $k^*$ correctly, that is,
$\tilde{k} \neq k^*$;
(iii) Alice finds a suitable pair, say $(k^*, \ell^*)$, but
but  $\tilde{\ell} \neq \ell^*$. We will analyse these events separately.
Consider the indicator random variable
\beq
\mathbf{J}(k,\ell) := \indI\left\{\mathbf{I}(k,\ell)=1 ~ \mbox{and}~ \alpha(k,\ell), \beta(k,\ell) > 1-4\eps_0\right\},
\enq
and consider the three events corresponding to the three sources of error identified above
\begin{eqnarray*}
\cE_1 &:=& \mbox{for all}~(k,\ell) \in \cC_1(1)\times\cC_2(1):\mathbf{J}(k,\ell)=0;\\
\cE_2 &:=&  \cE_1^c \mbox{ and } \tilde{k} \neq k^*; \\
\cE_3 &:=&  \cE_1^c \mbox{ and } \tilde{\ell} \neq \ell^*.
\end{eqnarray*}
\paragraph{Consider $\cE_1$:} We claim
\beq 
\label{event11a}
\Pr\{\cE_1\} \leq  2^{-r_1 - r_2 + I_{\infty}+ 2} + 2^{-r_1 + 4} + 2^{-r_2+4}.
\enq
We first show a lower bound on $\E\{\mathbf{J}(k,\ell)\}$. 
We observed in (P1) above that $\Pr\{\mathbf{I}(k,\ell)=1\} \geq
(1-\eps_\infty) 2^{-I_\infty}$. We account for (and exclude) the 
probability of the events $\alpha(k,\ell) \leq 1-4 \eps_0$
and $\beta(k,\ell) \leq 1- 4\eps_0$.
Let 
\[ \mathrm{Bad} = 
\{(u,v): \tr\left[\Lambda^B_u\rho^B_{f\left(u,v\right)}\right] \leq 1-4 \eps_0\}.
\]
We now upper bound $\displaystyle\Pr_{(U,V)}\{\mathrm{Bad}\}$ as follows.
\begin{align}
\Pr_{(U,V)}\{\mathrm{Bad}\} &= \Pr \left\{\left(1-\tr\left[\Lambda^B_U\rho^B_{f\left(U,V\right)}\right]\right) \geq 4\eps_0\right\}\nonumber\\
&\overset{a} \leq \frac{1-\E\left\{\tr\left[\Lambda^B_U\rho^B_{f\left(U,V\right)}\right]\right\}}{4\eps_0} \nonumber\\
\label{eq1}
&\overset{b}\leq \frac{1}{4},
\end{align}
where $a$ follows from Markov's inequality and $b$ follow from the definitions of $\Lambda^B_U$ and $\Gamma^{UB}$ and the fact that $\tr\left[\Gamma^{UB} \rho^{UB}\right] \geq 1-\eps_0 $. Thus,
\begin{align}\Pr\{\mathbf{I}(k,\ell)=1 \mbox{ and } \alpha(k,\ell) \leq 1-4\eps_0\}
  & = \sum_{(u,v) \in \mathrm{Bad}} p(u) p(v) \frac{p(u,v)}{2^{I_{\infty}}p(u) p(v)} \nonumber\\
  \label{eq2}
   &\leq \left(\frac{1}{4}\right) 2^{-I_{\infty}},\end{align}
 where the last inequality above follows from \eqref{eq1}.  
Similarly, \beq\label{eq3}\Pr\{\mathbf{I}(k,\ell)=1 \mbox{ and } \beta(k,\ell) \leq 1-4\eps_0\}
\leq \left(\frac{1}{4}\right) 2^{-I_{\infty}}.\enq We now lower bound  $\E\{\mathbf{J}(k,\ell)\}$ as follows.
\begin{align*}
\E\{\mathbf{J}(k,\ell)\} & = \Pr\left\{\mathbf{J}(k,\ell) =1 \right\}\\
&\geq \Pr\left\{\mathbf{I}(k,\ell) =1 \right\} - \Pr\{\mathbf{I}(k,\ell)=1 \mbox{ and } \alpha(k,\ell) \leq 1-4\eps_0\}- \\
&\hspace{5mm}\Pr\{\mathbf{I}(k,\ell)=1 \mbox{ and } \beta(k,\ell) \leq 1-4\eps_0\}\\
& \overset{a} \geq \left(1- \eps_{\infty} - \frac{1}{4} - \frac{1}{4}\right)2^{-I_\infty}\\
& \overset{b}               \geq 2^{-I_\infty - 2},
\end{align*}
where $a$ follows from property (P$1$) pertaining to $\mathbf{I}(k,\ell)$, \eqref{eq2} and \eqref{eq3} and $b$ follows because $\eps_\infty \leq \frac{1}{4}$.
Furthermore,
\[ \E\{\mathbf{J}(k,\ell) \mathbf{J}(k',\ell')\} \leq \E\{ \mathbf{I}(k,\ell) \mathbf{I}(k'\ell')\};\]
in particular, using properties (P2) and (P3) of $\mathbf{I}(k,\ell)$, we have
for $k' \neq k$ and $\ell' \neq \ell$,
\[ \E\{\mathbf{J}(k,\ell) \mathbf{J}(k',\ell)\}, \E\{\mathbf{J}(k,\ell) \mathbf{J}(k,\ell')\} \leq 2^{-2I_{\infty}}.\]
Also, $\mathbf{J}(k,\ell)$ and $\mathbf{J}(k',\ell')$ are independent whenever
$k\neq k'$ and $\ell\neq \ell'$. By Lemma~\ref{covering} (see Section~\ref{sec:mutualcovering}, set $\alpha \leftarrow \frac{1}{4}$, $q\leftarrow 2^{I_\infty}$),
\beq
\label{e111}
\Pr\{\cE_1\} \leq   2^{-r_1 -r_2 + I_\infty + 2} +
                       \frac{2^{r_1} + 2^{r_2}}{2^{r_1+r_2- 4}}
= 
2^{-r_1 - r_2 + I_{\infty}+ 2} + 2^{-r_1 + 4} + 2^{-r_2+4}.
\enq

From \eqref{e111} and our choice of the pair $(r_1, r_2)$ it now follows that
\beq
\label{event11}
\Pr\{\cE_1\} \leq 36\tilde{\eps}.
\enq
\paragraph{Consider $\cE_2$ and $\cE_3$:} 
We claim that 
\begin{align}
\label{errorB}
\Pr\{\cE_2\} &\leq
8\eps_0+2^{R_1+2r_1+r_2+2}2^{-I_{\infty}}2^{-I^B_0} \leq 8\eps_0 + 2\tilde{\eps};\\
\Pr\{\cE_3\}
& \leq
8\eps_0+2^{R_2+2r_2+r_1+2}2^{-I_{\infty}}2^{-I^C_0} \leq 8\eps_0+2\tilde{\eps}.
\label{errorC}
\end{align} 
To justify \eqref{errorB}, we have the following inequalities (below $k'$ 
takes ranges over $[2^{R_1+r_1}]$ and $(k,\ell)$ ranges over $\cC_1(1) \times \cC_2(1)$).
\begin{eqnarray}
\Pr\{\cE_1^c \mbox{ and }\tilde{k} \neq k^* \} 
&=&  
\E_{\cC}\left\{\mathbf{I}\{\cE_1^c\} 
\tr \left[
        \left(\mathbb{I}-\cT^B_{k^*}\right)
        \rho^B_{f\left(U[k^*], V[\ell^*]\right)}\right]\right\}\nonumber\\
&\overset{a} \leq &
2\mathbb{E}_{\cC}\left\{\mathbf{I}\{\cE_1^c\}\tr \left[\left(\mathbb{I}-\Lambda^B_{U[k^*]}\right)\rho^B_{f\left(U[k^*], V[\ell^*]\right)}\right]\right\}\nonumber\\
&& {} +\mathbb{E}_{\cC}\left\{4\sum_{k,\ell}\mathbf{I}\{k^*=k,\ell^*=\ell\}\sum_{k^\prime \neq k^*}\tr \left[\Lambda^B_{U[k^\prime]}\rho^B_{f\left(U[k^*], V[\ell^*]\right)}\right]\right\}\nonumber\\
&\overset{b}  \leq &  8\eps_{0} + 4\sum_{k,\ell,k^\prime \neq k}\mathbb{E}_{\cC}\left\{\mathbf{I}(k,l)\tr \left[\Lambda^B_{U[k^\prime]}\rho^B_{f\left(U[k], V[\ell]\right)}\right]\right\}\nonumber\\
&\overset{c} = &  8\eps_0 + 4\sum_{k,\ell,k^\prime \neq k}\sum_{u,v,u^\prime}2^{-I_{\infty}}P_{U}(u)P_{V}(v)\frac{P_{UV}(u,v)}{P_{U}(u)P_{V}(v)}P_{U}(u^\prime)\tr\left[\Lambda^B_{u^\prime}\rho^B_{f\left(u,v\right)}\right] \nonumber\\
&\leq &  8\eps_0 +2^{r_1+r_2+R_1+r_1+2}2^{-I_{\infty}}\tr\left[\Gamma^{UB}\left(\rho^U\otimes\rho^B\right)\right]\nonumber\\
&\overset{d}\leq & 8\eps_0+2^{R_1+2r_1+r_2+2}2^{-I_{\infty}}2^{-I^B_0},\nonumber\\
\label{error1}
&\overset{e}\leq & 8\eps_0+2\tilde{\eps}, \nonumber
\end{eqnarray}
where $a$ follows from the Hayashi-Nagaoka operator inequality \cite{Hayashi-noniid}; $b$
follows from the definition of the event $\cE_1$ and because our encoding 
ensures that $\alpha(k^*,\ell^*) > 1 - 4\eps_0$
; $c$ follows from the
definition of $\mathbf{I}(k,l)$; $d$ follows from the definition of
$\rho^U$, $\rho^B$ and $\Gamma^{UB}$; $e$ follows because $R_1$, $r_1$ and $r_2$ satisfy
(\ref{eq:rfirstq})--(\ref{eq:r2firstq}). Similarly, we justify (\ref{errorC}).
Thus, from \eqref{event11}, \eqref{errorB} and \eqref{errorC} it follows that
\beq
\Pr\left\{\tilde{M}_1\neq1 \cup \tilde{M}_2 \neq 1\right\} \leq 40\tilde{\eps}+16\eps_0. \nonumber
\enq
The above upper bound on the probability of error applies to every
$(m_1,m_2)$ as we average over the choices of the codebook; by
linearity of expectation, this upper bound holds in expectation
when $(m_1,m_2)$ is chosen randomly. It follows that there is a
fixed codebook for which the expected error is bounded by 
$40\tilde{\eps} + 16 \eps_0$. This completes the proof.

\section{The classical one-shot bound}
\label{sec:classical}

The proofs in this section are just translations of the proof for the
quantum case presented above; we reproduce the common parts for the
sake of completeness.
\begin{definition}
A classical broadcast channel consists of an input alphabet $\cX$,
two output alphabets $\cY$ and $\cZ$ and the probability transition
function $p_{YZ\mid X}$.
\end{definition}
\begin{definition}
An $(R_1,R_2, \eps)$-code for a classical broadcast channel $C =
\{p_{YZ \mid X}\}$ consists of
\begin{itemize}
\item an encoding function $F: [2^{R_1}] \times [2^{R_2}] \rightarrow
  \cX$, and
\item two decoding functions $D_1: \cY \rightarrow [2^{R_1}]$ and 
$D_2: \cZ \rightarrow [2^{R_2}]$ 
\end{itemize}
such that
\[ \Pr\{(M_1,M_2) \neq (D_1(Y), D_2(Z))\} \leq \eps, \]
where $(M_1,M_2)$ are uniformly distributed over 
$[2^{R_1}] \times [2^{R_2}]$, and $Y$ and $Z$ satisfy 
$\Pr\{(Y=y, Z=z) \mid  M_1 = m_1, M_2=m_2\} = p_{YZ|X}(yz \mid F(m_1,m_2))$.
\end{definition}
\subsection{One-shot Marton inner bound for the classical broadcast channel}
Our one-shot version of the Marton inner bound will be stated
in terms of min and max \renyi divergences which are defined
as follows.
\begin{definition}(Smooth classical Min \renyi divergence \cite{renner-isit-2009})\\
\label{mindiv}
For random variables $(U,V)\sim p_{UV}$ with range  $\cU\times\cV$
and $\eps\in [0,1)$ we have the following.
\beq
I^{\eps}_{0}[U;V]:=\sup_{\substack{\cA\subseteq \cU\times\cV\\ p_{UV}(\cA)\geq 1-\eps}}-\log \sum_{(u,v) \in \cA}p_{U}(u)p_{V}(v). \nonumber
\enq
\end{definition}

\subsection{Code generation}
We need to describe a function $F: [2^{R_1}] \times [2^{R_2}]
\rightarrow \cX$. We will adapt the scheme suggested by Marton as
presented in El Gamal and Kim~\cite{Gamal-Kim-book}, to the one-shot setting.

\paragraph{The random codebook:} Let $(U,V,Y,Z)$ and $f$ be as in the
statement of the theorem, and $(R_1,R_2)$ satisfy the required
inequalities.  In the following we set
\begin{eqnarray*}
I_\infty  & = & I^{\eps_\infty}_\infty[U;V]; \\
I_0^B     & = & I^{\eps_0}_0[U;Y];\\
I_0^C     & = & I^{\eps_0}_0[V;Z].\\
\end{eqnarray*}
Let $\cA_1$ be the set in the definition of $I_0^B =
I_0^{\eps_0}(U;Y)$ such that $p_{UY}(\cA_1) > 1 -
\eps_0$ and $\sum_{(u,y) \in \cA_1} p_U(u) p_Y(y) \leq
2^{-I_0^B}$. Similarly, let $\cA_2$ be the set such that
$p_{VZ}(\cA_2) > 1 - \eps_0$ and $\sum_{(v,z) \in \cA_1} p_V(v)
p_Z(z) \leq 2^{-I_0^C}$. Choose positive integers $r_1$ and $r_2$ such that
\newcommand{\slack}{4 \log \frac{1}{\tilde{\eps}} -1}
\begin{eqnarray}
 R_1 + r_1 &\leq& {I_0^B - \slack}; \label{eq:rfirst} \\
 R_2 + r_2 &\leq& {I_0^C - \slack}; \\
r_1, r_2 & \geq & \log \frac{1}{\thirdepsilon}; \\
r_1 + r_2 & = & \ceil{I _\infty + 3\log \frac{1}{\thirdepsilon}}. 
\label{eq:rlast}
\end{eqnarray}
[The existence of such $r_1$ and $r_2$ was justified in the quantum case.]

\begin{comment}
\label{rem:overcounting}
The main difference from the usual calculation is in the
(\ref{eq:rlast}).  One usually imposes a lower bound on $r_1+r_2$ in
order to ensure the {\em covering property} that with high probability
there is a codeword available in the intersection of the two message
bands (see $\event_1$ below). The value for $r_1+r_2$ set in
(\ref{eq:rlast}) suffices to ensure that such a code word is
available with high probability. However, by insisting that $r_1+r_2$
not exceed the value by too much (we require equality in
(\ref{eq:rlast})), we ensure that there are not too many such
codewords to choose from, which intuitively makes the decoding
unambiguous.
\end{comment}
Let $U[1]$, $U[2]$,$\ldots, U[2^{R_1+r_1}]$ be drawn independently
according to the distribution of $U$; similarly, let $V[1]$,
$V[2]$,$\ldots, V[2^{R_2+r_2}]$ be drawn according the distribution of
$V$. These samples will be associated with rows and columns of a
$2^{R_1+r_1} \times 2^{R_2+r_2}$ matrix $\cC$, whose entries will be
elements of $\cX \cup \{\star\}$. The entry $\cC[k,\ell]$ will be
determined as follows.

For each pair $(k,\ell)$, let $\eta(k,\ell)$ be chosen independently and
uniformly from $[0,1]$. Let $\rvI(k,\ell)$ be the $0$-$1$ indicator
random variable defined by
\[ \rvI(k,\ell) = \indI\left\{ \eta(k,\ell) \leq
\frac{p(U[k],V[\ell])}{2^{I_{\infty}}{ p(U[k]) p(V[\ell])}} \right\}.\]
Then, $\cC[k,\ell] = f(U[k], V[\ell])$ if $\rvI(k,\ell) = 1$, and
$\cC[k,\ell]=\star$ otherwise. Thus, $\cC$ is a random matrix of entries,
determined by the random choices of $(U[k], V[\ell], \eta(k,\ell))$
for $k=1,2,\ldots, 2^{R_1+r_1}$ and $\ell = 1,2, \ldots, 2^{R_2+r_2}$;
we will call this (the random matrix, together with all the associated
random choices $U[k]$, $V[\ell]$ and $\eta(k,\ell)$) the random
codebook $\cC$. Later we will fix one realization of $\cC$.

Our encoding function $F: [2^{R_1}] \times [2^{R_2}] \rightarrow \cX$ will be
based on $\cC$. We partition the row indices of $\cC$ into $2^{R_1}$
classes each with $2^{r_1}$ elements; let the $i$-th class $\cC_1(i) =
\{(i-1)2^{r_1}+1, (i-1)2^{r_1}+ 2, \ldots, i 2^{r_1}\}$. Similarly, we
partition the column indices into $2^{R_2}$ classes, where the $j$-th
class $\cC_2(j) = \{(j-1)2^{r_2}+1, (j-1)2^{r_2}+ 2, \ldots, j
2^{r_2}\}$.  If $\cC[i,j] = x \neq \star$, then let
\begin{eqnarray*}
\alpha(i,j) &=& \sum_{y: (U[i], y) \in \cA_1} p(y \mid x);\\
\beta(i,j) &=&  \sum_{z: (V[j], z) \in \cA_2} p(z \mid x);
\end{eqnarray*}
if $\cC[i,j]=\star$, let $\alpha(i,j), \beta(i,j)= - \infty$.  For a
pair of messages $(m_1, m_2)$, let $F(m_1,m_2) = \cC[i,j]$, where $(i,j)
\in \cC_1(m_1) \times \cC_2(m_2)$ is the lexicographically first pair such
that $\alpha(i,j), \beta)({i,j}) > 1 - 4\eps_0$; if no such $(i,j)$
exists, then let $F(m_1,m_2)$ be the first element of $\cX$.

\paragraph{Joint typicality versus rejection sampling:} In the standard
argument~\cite{Gamal-Kim-book}, the indicator random variable
$\rvI(k,\ell)$ stands for joint typicality of $U[k]$ and
$V[\ell]$. The rejection sampling based on $I_\infty$, has the same
effect. We list below its properties.
\begin{enumerate}
\item $\E\{\rvI(k,\ell)\} \geq (1-\eps_{\infty}) 2^{-I_\infty}$.
\item For all $u$ and $v$,
\begin{eqnarray}
\E\{\rvI(k,\ell) \mid U[k]=u\} 
&\leq & \sum_v p(v) \frac{p(u,v)}{2^{I_\infty} p(u)p(v)} \leq 2^{-I_\infty}; \\
\E\{\rvI(k,\ell) \mid V[\ell]=v\}  & \leq & 
\sum_u p(u) \cdot \frac{p(u,v)}{2^{I_\infty} p(u)p(v)} \leq
2^{-I_\infty};\\
\E\{\rvI(k,\ell)\} & \leq & 2^{-I_\infty}.
\end{eqnarray}
\item If $\ell \neq \ell'$, then $\rvI(k,\ell)$ and $\rvI(k,\ell')$
  are conditionally independent given $U[k]$; if $k \neq k'$, then
  $\rvI(k,\ell)$ and $\rvI(k',\ell)$ are conditionally independent
  given $V[\ell]$.

\item If $k \neq k'$ and $\ell \neq \ell'$, then 
 $\rvI(k,\ell)$ and $\rvI(k',\ell)$ are independent.
\end{enumerate}

\paragraph{Decoding:} We first consider Bob's strategy for recovering
$m_1$ on receiving the channel output $\ty$: let $D_1(\ty)$ be the
smallest $\tm_1$ such that there is a $k \in \cC_{1}({\tm_1})$ such that
$(U[k],\ty) \in \cA_1$.  Similarly, Charlie's strategy is determined
using the set $\cA_2$: $D_2(\tz)$ is the smallest $\tm_2$ such that
there is an $\ell \in \cC_2({\tm_2})$ such that $(V[\ell],\tz) \in
\cA_2$. In both cases, if an appropriate $U[k]$ or $V[\ell]$ is not
found, the answer $1$ is returned.  (In fact, we will show that whp
there is a unique such $(k,\ell)$ in $[2^{R_1+r_1}]$; if there
is a pair $(k,\ell)$ satisfying the above requirements, but different from the one used by Alice, then we will consider it as an error.)

\subsection{Proof of Theorem~\ref{thm:classical}}
Suppose a pair of messages $(m_1,m_2) \in [2^{R_1}] \times
[2^{R_2}]$ is transmitted by Alice using the above scheme and is
decoded as $(\tm_1, \tm_2)$ by Bob and Charlie.  We wish to show that
the probability (averaged over the choice of the codebook) that
$(\tm_1,\tm_2) \neq (m_1,m_2)$ is at most $\eps$.  By the symmetry
in the generation of the code book, it is enough to prove this claim
for $(m_1,m_2)=(1,1)$.

We identify three sources of error. First, we regard as error those
cases where there is no pair $(k,\ell) \in \cC_1(1) \times \cC_2(1)$ for
which $\rvI(k,\ell) =1$ and $\alpha(i,j), \beta(i,j) > 1
-4\eps_0$; second, it may happen that even though such a pair $(k^*,
\ell^*)$ is found, we have $(U[k^*],\ty) \not \in \cA_1$ (here, as
before, $\ty$ refers to the channel output received by Bob) or
$(V[\ell^*], \tz) \not \in \cA_2$; third, Alice or Bob may not recover
$(k,\ell)$ uniquely, for it may happen that $(U[k'],\ty) \in
\cA_1$, for some $k' \neq k$ or $(V[\ell'],\tz) \in \cA_2$, for some
$\ell' \neq \ell$. Consider the indicator random variable
\[ \rvJ(k,\ell) = \indI\{ \rvI(k,\ell)=1 \mbox{ and } \alpha(k,\ell), \beta(k,\ell) > 1- 4 \eps_0\},\]
and the events
\begin{eqnarray*}
\event_1 & = & \mbox{for all $(k,\ell) \in \cC_1(1) \times \cC_2(1): 
\rvJ(k,\ell)=0$} ;\\
\event_{2,B} & = & \event_1^c \mbox{ and } (U[k^*],\ty) \not \in \cA_1;\\
\event_{2,C} & = & \event_1^c \mbox{ and } (V[\ell^*],\tz) \not \in \cA_2;\\
\event_{3,B} & = & \event_1^c \mbox{ and } 
\mbox{for some $k' \neq k^*: (U[k'],\ty) \in \cA_1$}; \\
\event_{3,C} & = & \event_1^c \mbox{ and } 
\mbox{for some $\ell' \neq \ell^*: (V[\ell'],\tz) \in \cA_2$}.
\end{eqnarray*}
Clearly, if we exclude all the above events, then Bob and Charlie
indeed recover the pair $(k^*, \ell^*)$ used by Alice.
%
%
\begin{claim}
\begin{enumerate}
\item[(i)] $\Pr\{\event_1\} \leq  2^{-r_1 - r_1 + I_{\infty}+ 2} +
                           2^{-r_1 + 4} + 2^{-r_2+4} \leq 36\tilde{\eps}$.
\item[(ii)] $\Pr\{\event_{2,B}\}, \Pr\{\event_{2,C}\} \leq 4 \zeroepsilon$;
\item[(iii)] $\Pr\{\event_{3,B}\} \leq 2^{2r_1+r_2+R_1} 2^{-I_\infty} 2^{-I_0^B}\leq \frac{\tilde{\eps}}{2}$ and
      $\Pr\{\event_{3,C}\} \leq 2^{r_1+2r_2+R_2} 2^{-I_\infty} 2^{-I_0^C}\leq\frac{\tilde{\eps}}{2}$.
\end{enumerate}
\end{claim}
Then, from our claim
and the union bound, we conclude
\[
\Pr\{\mbox{error} \}  \leq 37\tilde{\eps}+8\eps_0.
\]
The above upper bound on the probability of error applies to every
$(m_1,m_2)$ as we average over the choices of the codebook; by
linearity of expectation, this upper bound holds in expectation
when $(m_1,m_2)$ is chosen randomly. It follows that there is a
fixed codebook for which the expected error is bounded by 
$37\thirdepsilon + 8 \eps_0$.

It remains to establish the claim above.
\paragraph{Consider $\event_1$:} Since $\Pr\left\{\alpha(k,\ell)
  \leq (1-4 \eps_0)\right\}, \Pr\left\{\beta(k,\ell) \leq (1-4 \zeroepsilon)\right\}
\leq \frac{1}{4}$ and $\eps_\infty \leq \frac{1}{4}$, we have
\[
\E\{\rvJ(k,\ell)=1\} \geq \left(\frac{1}{2}-\eps_\infty\right)2^{-I_\infty}
                \geq 2^{-I_\infty - 2}. 
\]
Furthermore,
\[ \E\{\rvJ(k,\ell) \rvJ(k',\ell')\} \leq \E\{ \mathbf{I}(k,\ell) \mathbf{I}(k',\ell')\};\]
in particular, for $k' \neq k$ and $\ell' \neq \ell$,
\[ \E\{\rvJ(k,\ell) \rvJ(k',\ell)\}, \E\{\rvJ(k,\ell) \rvJ(k,\ell')\} \leq 2^{-2I_{\infty}}.\]
Also, $\rvJ(k,\ell)$ and $\rvJ(k',\ell')$ are independent whenever
$k\neq k'$ and $\ell\neq \ell'$. By Lemma~\ref{covering},
\[
\Pr\left\{\event_1\right\} \leq   2^{-r_1 -r_2 + I_\infty +2} +
                       \frac{2^{r_1} + 2^{r_2}}{2^{r_1+r_2-4}}.
\]
Thus, from our choice of $r_1$ and $r_2$ it now easily follows that
$\Pr[\event_1]\leq 36\tilde{\eps}$.
\paragraph{Consider $\event_{2,B}$, $\event_{2,C}$:} It follows immediately from the definition of $\event_1$, that 
\[ \Pr\left\{\event_{2,B}\right\}, \Pr\left\{\event_{2,C}\right\} \leq 4\eps_0.\]

\paragraph{Consider $\event_{3,B}$, $\event_{3,C}$:} We will focus on $\event_{3,B}$;  
similar arguments are applicable to $\event_{3,C}$ Fix a codebook. For
$(k,\ell) \in \cC_1(1) \times \cC_2(1)$ such that $\rvI(k,\ell) = 1$,
let
\begin{eqnarray*}
\Gamma(k,\ell) & = & \{y: (U(k'),y) \in \cA_1 \mbox{ for some } k' \neq k\};\\
\gamma(k,\ell) & = & \sum_{y \in \Gamma(k,\ell)} p(y \mid \cC[k,\ell]) \\
& \leq & \sum_{k' \neq k} \sum_{y: (U(k'),y) \in \cA_1} p(y \mid \cC[k,\ell]).
\end{eqnarray*}
Then, (here $(k,\ell)$ ranges over $\cC_1(1) \times \cC_2(1)$)
\begin{eqnarray*}
\Pr\{\event_{3,B} \} 
&\leq& \sum_{k,\ell} \indI\{k^*=k,\ell^*=\ell\} \gamma(k,\ell)\\
&\leq& \sum_{k,\ell} \mathbf{I}[k,\ell] \gamma(k,\ell).
\end{eqnarray*}                   
[Note the last inequality involves over counting; we can afford it 
because of the upper bound on $r_1+r_2$ imposed through (\ref{eq:rlast}).]
Now, averaging over all code books, we have (below $k'$ 
takes ranges over $[2^{R_1+r_1}]$ and $(k,\ell)$ ranges over $\cC_1(1) \times \cC_2(1)$)
\begin{eqnarray}
\Pr\{\event_{3,B}\} 
& \leq &  \sum_{k,\ell} \E\{\mathbf{I}[k,\ell] \gamma(k,\ell) \}\nonumber\\
& \leq & 
\sum_{k,\ell,k' \neq k} \sum_{u,v,u'} \sum_{y: (u',y) \in \cA_1}
p(u) p(v) \left(\frac{p(u,v)}{2^{I_\infty} p(u)p(v)}\right) p(u') p(y | f(u,v)) \nonumber\\
& \leq &
\sum_{k,\ell,k' \neq k} \sum_{u,v,u'} \sum_{y: (u',y) \in \cA_1}
2^{-I_\infty} p(u,v) p(u') p(y \mid f(u,v)) \nonumber\\
& \leq &
\sum_{k,\ell, k' \neq k}  \sum_{(u',y) \in \cA_1} 
\left( \sum_{u,v} 2^{-I_\infty} p(u,v) p(y \mid f(u,v)) \right) p(u') \nonumber\\
& \leq &
\sum_{k,\ell,k' \neq k}  2^{-I_\infty} \left( \sum_{(u',y) \in \cA_1}  p(y) p(u') \right)\nonumber\\
& \leq &
\sum_{k,\ell,k' \neq k}  2^{-I_\infty} 2^{-I_0}\nonumber\\
&\leq & 2^{r_1+r_2+R_1 + r_1} 2^{-I_\infty} 2^{-I_0}.
\label{e3b}
\end{eqnarray}
Thus, from \eqref{e3b} and by our choice of $R_1$, $r_1$ and $r_2$ as mentioned in (\ref{eq:rfirst})--(\ref{eq:rlast}) the desired result follows, i.e.,
\beq
\Pr\{\event_{3,B}\} \leq \frac{\tilde{\eps}}{2}. \nonumber
\enq

\section{Existence of good codewords}
In this section we prove a lemma which helps to prove the existence of good codewords. Such lemma is refereed to as mutual covering lemma in the information literature \cite[Lemma 8.1]{Gamal-Kim-book}. The claim of this lemma follows straightforwardly from Chebyshev's inequality.
\label{sec:mutualcovering}
\begin{lemma}
\label{covering}
Suppose $0 < q \leq  1$. Let $Z = \sum_{k=1}^{r} \sum_{\ell=1}^{s}
\rvJ(k,\ell)$, where the 0-1 random variables $\rvJ(k,\ell)
\in \{0,1\}$ are such that
\begin{eqnarray*}
\E\{\rvJ(k,\ell)\} & \geq & \alpha q \, ;\\
\E\{\rvJ(k,\ell)\rvJ(k,\ell')\} & \leq & q^2 \ \mbox{whenever $\ell \neq \ell'$} \ ;\\
\E\{\rvJ(k,\ell)\rvJ(k',\ell)\} & \leq & q^2 \ \mbox{whenever $k \neq k'$} \ ;
\end{eqnarray*}
furthermore, $\rvJ(k,\ell)$ and $\rvJ(k',\ell')$ are independent
whenever $k\neq k'$ and $\ell\neq \ell'$.
Then,
\[ \Pr\{ Z=0 \} \leq \frac{1}{\alpha rsq} + \frac{r+s}{\alpha^2 rs}.\]
\end{lemma}
\begin{proof}
We will use Chebyshev's inequality. 
We have
\begin{eqnarray}
\E\{Z\}  & \geq & \alpha rs q\, ;\\
\vr\{Z\} & = &  \E\{Z^2\} - \E\{Z^2\} \\
 & \leq & \sum_{(k,\ell), (k'\ell')} 
\left( \E\{\rvJ(k,\ell) \rvJ(k',\ell')\} - \E\{\rvJ(k,\ell)\} \E\{\rvJ(k'\ell')\} 
\right)
\\
& \leq & \E\{Z\} + rs (r+s) q^2,
\end{eqnarray}
where we used the fact that 
$\E\{\rvJ(k,\ell) \rvJ(k',\ell')\} - \E\{\rvJ(k,\ell)\} \E\{\rv(k'\ell')\}$
whenever $k\neq k'$ and $\ell\neq \ell'$.
Then, by Chebyshev's inequality, we have
\begin{eqnarray*}
\label{cheby}
\Pr\{Z=0\} & \leq & \frac{\vr\{Z\}}{\E\{Z\}^2} \\
           & \leq & \frac{\E\{Z\} + rs(r+s)q^2}{\E\{Z\}^2} \\
           & \leq & \frac{1}{\alpha rsq} + \frac{r+s}{\alpha^2 rs}.
\end{eqnarray*}
This completes the proof.

\end{proof}

\section{Asymptotics}
\label{Asymptotics}
As stated in the introduction our analysis immediately implies similar
bounds in the asymptotic iid and information spectrum settings. In this
section, we formally verify these claims for appropriate
classical-quantum channels; similar bounds also follow in the
classical setting, but we do not discuss them separately. 

Suppose we are given a sequence $\vec{\cH} = \left\{\cH^{(n)}\right\}_{n=1}^{\infty}$ of Hilbert spaces and a sequence $\vec{\cN} = \left\{\cN^{(n)}\right\}_{n=1}^{\infty}$ of channels $\cN^{(n)} : \cX^n \to \mathcal{S} (\cH^{\otimes n}_A \otimes \cH^{\otimes n}_B)$. An important example is the iid setting when $\cN^{(n)}(X^n) := \cN(X_1)\otimes \cN(X_2)\otimes \cdots \otimes \cN(X_n)$ for $X^n := (X_1, X_2, \cdots, X_n)$, where we assume that each coordinate of the sequence $X^n$ are independent and identically distributed. When we do not make such assumptions then we call this extremely general approach as information spectrum (non-iid) approach \cite{Hayashi-noniid}. An asymptotically achievable rate pair $(R_1, R_2)$ is then defined as follows. 
\begin{definition}
\label{martasym}
A rate pair $(R_1,R_2)$ is asymptotically achievable for a sequence of channel ${\vec{\cN}} = \left\{\cN^{(n)}\right\}_{n=1}^{\infty}$ if and only if there exists an encoding function $F_{n}$, where $F_{n}: [2^{R^{(n)}_1}] \times [2^{R^{(n)}_1}] \to \cX^n$ and a pair of decoding POVMs $\{\cT^{B^{(n)}}_{m_1}: m_1 \in [2^{R^{(n)}_1}]\}$ and $\{\cT^{C^{(n)}}_{m_1}: m_2 \in [2^{R^{(n)}_2}]\}$ such that
\begin{align*}
R_1 &\leq \liminf_{n \to \infty} \frac{R^{(n)}_1}{n}\\
R_2 & \leq \liminf_{n \to \infty} \frac{R^{(n)}_2}{n}
\end{align*}
and $\lim_{n\to \infty}\frac{1}{2^{R^{(n)}_1+R^{(n)}_2}}\sum_{(m_1,m_2) \in [2^{R^{(n)}_1}] \times [2^{R^{(n)}_2}]}
\tr\left[\left(\mathbb{I} - \cT^{B^{(n)}}_{m_1} \otimes \cT^{C^{(n)}}_{m_2}\right)
      \cN^{(n)}(F(m_1,m_2))\right]= 0$.
\end{definition}

\paragraph{Asymptotic iid setting:} The bound derived by Savov and 
Wilde~\cite{savov-wilde-arxiv-v3} in the iid setting, which was restated
as Theorem~\ref{savov-wilde} in the introduction, follows from Definition \ref{martasym} and
Theorem~\ref{martonquantum-oneshot} because of the following convergence results.

\begin{theorem} 
\begin{description}
\item[$(a)$]{(Ogawa and Nagaoka \cite{ogawa-nagaoka-2000-strong})
Let $\rho^{UB}$ be a classical-quantum state, and let $\rho^{U^nB^n}$ be its
$n$-fold tensor. Then, for all $\eps > 0$, we have
\begin{equation}
I[U;B] =  \lim_{n \rightarrow \infty} \frac{1}{n}I_0^{\eps}[U^n;B^n],
\end{equation}
where $I[U;B]$ is computed with respect to the state $\rho^{UB}$ and
$I^{\eps}_{0}[U^n;B^n]$ is computed with respect to the state $\rho^{U^nB^n}$.}
 \item[$(b)$]{(Datta \cite{datta-2009-relative})
For a pair of classical random variables $(U,V)$, let
represent $(U^n,V^n)$ be $n$ independent copies of $(U,V)$. Then, for all $\eps>0$, we have
\begin{equation}
I[U;V] =  \lim_{n \rightarrow \infty} \frac{1}{n} I_\infty^{\eps}[U^n;V^n].
\end{equation}}
\end{description}
\end{theorem}
\begin{comment}
Note that though Theorem~\ref{savov-wilde} was formulated
in~\cite{savov-wilde-isit}; its complete justification appeared later
in~\cite{savov-wilde-arxiv-v3}; their analysis which works directly in
the asymptotic setting makes crucial use of the over counting argument and some analysis techniques
that first appeared in the preliminary version of our work
\cite{radhakrishnan-sen-warsi-archive-v1}.
\end{comment}

\paragraph{Asymptotic non-iid setting:}  We first review the basic
definitions in the asymptotic non-iid setting and formulate the rate
region. In this setting, we again have an infinite sequence of states
(with respect to which the asymptotic analysis is performed), but
successive states will not be obtained by independent repetitions of a
basic state. The analogs of the quantities $I^{\eps_0}[U;B]$ and $
I^{\eps}_{\infty}[U;V]$ in this setting are as follows.

Let $\{U^n\}_{n=1}^\infty$ be a sequence of random variables, where $U^n$
takes values in ${\cU}^n$. Furthermore, for each $n$ and each $u^n
\in {\cal U}^n$, let $\rho^{B^n}_{u^n}$ be a quantum state in the Hilbert
space ${\cal H}_n$. Let 
$\bm{\rho^{UB}}:=\left\{\rho^{U^nB^n}\right\}_{n=1}^{\infty}$ be a
sequence of classical-quantum states, where 
\beq
\rho^{U^nB^n}:=\sum_{u^n \in
  \cU^n}p_{U^n}(u^n)\ket{u^n}\bra{u^n}^{u^n}\otimes \rho^{B^n}_{u^n}. 
\enq
With respect to this sequence $\underline{{I}} [\mathbf{U};\mathbf{B}]$ is
defined as follows.
\begin{definition}(Spectral inf quantum mutual information rate \cite{datta-byeondiid-2006})
The spectral inf mutual information rate for $\rho^{UB}$ is 
\beq
\underline{{I}} [\mathbf{U};\mathbf{B}]:= \sup\left\{\gamma: \lim_{n\to \infty}\sum_{u^n \in \cU^n}p_{U^n}(u^n)\tr \left[\left\{\rho^{B^n}_{u^n}\succeq 2^{n\gamma}\rho^{B_{n}}\right\}\rho^{B^{n}}\right]=1\right\}, \nonumber
\enq
where $\rho^{B^{n}}=\tr_{U^n}\left[\rho^{U^nB^n}\right]$ and $\left\{\rho^{B^n}_{u^n}\succeq 2^{n\gamma}\rho^{B^{n}}\right\}$ is the projector onto the positive Eigen space of the operator $\rho^{B^n}_{u^n}-2^{n\gamma}\rho^{B_{n}}$.
\end{definition}

Let ${(\bf{U},\bf{V})}:=\left\{(U^n,V^n)\right\}_{n=1}^{\infty}$ be a
sequence of pairs of random variables where $(U^n,V^n)$ take values
in $\cU^n \times \cV^n$. With respect to this sequence $\overline{{I}} [\mathbf{U};\mathbf{V}]$ is defined as follows.
\begin{definition}(Spectral sup classical mutual information rate \cite{han-book})
 The spectral sup classical mutual information rate between ${\bf{U}}$
 and ${\bf{V}}$ is 
\beq \overline{I}{[\bf{U};\bf{V}]}:=
 \inf\left\{\lambda: \lim_{n\to\infty}\Pr\left\{
 \frac{1}{n}\log\frac{p_{U^nV^n}}{p_{U^n}p_{V^n}}>\lambda\right\}=0\right\},
 \nonumber 
\enq 
where the probability is calculated with respect to $p_{U^nV^n}$.
\end{definition}

With this, we may formulate the Marton inner bound in the information
spectrum setting as follows. 

\begin{theorem}
Let $\left\{\cN^{(n)}(x^n):=\rho^{B^nC^n}_{x^n}\right\}_{n=1}^{\infty}$ be a sequence of general classical-quantum broadcast channel where for every $n$ and $x^n$, $\rho^{B^nC^n}_{x^n} \in \mathcal{S}\left(\cH_B^{\otimes n} \otimes \cH^{\otimes n}_{C}\right)$. Let $\{F_n\}_{n=1}^{\infty}$ be a sequence of functions where for every $n$, $f_n: \cU^n\times\cV^n \to \cX^n$; consider the sate 
\beq
\rho^{U^nV^nB^nC^n}=\sum_{(U^n,V^n)\in (\cU^n\times\cV^n)}p_{U^nV^n}\ket{u^n}\bra{u^n}^{U^n}\otimes\ket{v^n}\bra{v^n}^{V^n}\otimes\rho^{B^nC^n}_{f_n(u^n,v^n)} \nonumber.
\enq
Let $(R_1,R_2)$ be such that 
\begin{align}
\label{s1}
R_1 &< \underline{{I}} [\mathbf{U};\mathbf{B}]\\
\label{s2}
R_2 &< \underline{{I}} [\mathbf{V};\mathbf{C}]\\
\label{s1+s2}
R_1+R_2 & < \underline{{I}} [\mathbf{U};\mathbf{B}]+\underline{{I}} [\mathbf{V};\mathbf{C}]-\overline{I}{[\bf{U};\bf{V}]}.
\end{align}
Then, $(R_1,R_2)$ is achievable. The information theoretic quantities mentioned in \eqref{s1}, \eqref{s2} and \eqref{s1+s2} are calculated with respect to the sequence of states $\left\{\rho^{U^nV^nB^nC^n}\right\}_{n=1}^\infty$.

\end{theorem}
\begin{proof}
The proof immediately follows from Theorem \ref{martonquantum-oneshot}, Definition \ref{martasym} and from the observation that for every $\eps \in (0,1), \gamma < \underline{{I}} [\mathbf{U};\mathbf{B}], \lambda>\overline{I}{[\bf{U};\bf{V}]}$ and for $n$ large enough we have
\begin{align*}
\frac{1}{n}I^{\eps}_0[U^n;B^n] &\geq \gamma\\
\frac{1}{n}I^{\eps}_{\infty}[U^n;V^n] &\leq \lambda.
\end{align*}
\end{proof}
\begin{comment} An important variation of the problem discussed in this section is that of Marton inner bound with common message. In this case Alice wants to transmit a message triplet $(M_0,M_1, M_2)$. As before $M_1$ is meant for Bob and $M_2$ is meant for Charlie. However, the message $M_0$ is meant for both Bob and Charlie and is called as common message. We note here that the techniques developed in this chapter only deals with the case when there is no common message. 

\end{comment}
\subsection*{Acknowledgments}

We are grateful to Vinod Prabhakaran and Mark Wilde for useful discussions and comments.

\bibliographystyle{ieeetr}
\bibliography{master}
\end{document}